# Two Vocabularies, One Phenomenon: Metadata Bias in AI Evidence Synthesis on Fertility Decline

Declining fertility is one of the defining policy questions of the next decade, and increasingly, what policymakers know about it is shaped by AI-synthesising the evidence base. But ask such a tool about reproduction and the answer depends on the word you use. The same phenomenon, framed clinically (e.g, infertility, IVF) or socially (e.g., childlessness, fertility intentions), is catalogued with radically different completeness. And the catalogue as much if not more than the underlying scholarship, is what AI synthesis begins with (Bolaños et al., 2024). That databases under-index the social sciences, books, and grey literature is well established (Visser et al., 2021). What is new here is holding the topic fixed and asking whether metadata gaps act as a hidden policy filter on a single contested issue: the determinants of (in)fertility.

We use two OpenAlex queries on the same phenomenon: a clinical basket (infertility, subfertility, ART, IVF, fecundity; n=101,645) and a social basket (childlessness, social infertility, fertility intentions, reproductive decision-making; n=3,646). We compare them on metadata completeness, open access, output type, and institutional provenance. The social framing is consistently less machine-legible: output skewed to books and dissertations, authorship university- rather than healthcare-based. Open access rates are essentially equal (43.1% vs 41.3%), so the gap is in indexing depth, not paywalls, suggesting simple OA mandates will not fix it. On this same mixed literature, even before any coverage bias enters the picture, LLM tools already miss more than they catch when asked to extract hypotheses and claims (Uprety et al., 2025); the bias documented here compounds an already-imperfect extraction stage.

The governance implication is direct: discoverability infrastructure should be audited alongside models, prompts, and inclusion criteria. Metadata gaps are not housekeeping; they are an equity question about what counts as evidence.

## 1. Introduction

In 2023 the American Society for Reproductive Medicine redefined infertility as a need for medical intervention rather than biological failure (ASRM, 2023), extending the clinical category into terrain that social-science scholarship had long studied under different vocabulary (e.g., desire, partnership, access). The two framings now describe overlapping phenomena from incompatible starting points, and which framing reaches policy depends on AI-assisted synthesis tools that mediate the search, selection, and presentation of evidence (Sousa et al., 2026).

This paper is organised around two interrelated research questions on discoverability and its implications for governance. RQ1: Is the social-science evidence on fertility-determinants as machine-findable as the medical evidence, across indexing depth, identifier coverage, and full-text availability? RQ2: Should discoverability infrastructure be treated as a governance variable for AI evidence synthesis, audited alongside models, prompts, and inclusion criteria?

## 2. Data and Research Methodology

Mapping how this discourse has evolved, from clinical success rates to family-enabling policies, requires seeing both vocabularies clearly. We operationalise the two framings through controlled title-only queries against OpenAlex conducted on 30/05/26, chosen to maximise inclusivity across output types, including those without abstracts:

- *[CLINICAL] "*infertility" OR "subfertility" OR "assisted reproductive technology" OR "in vitro fertilization" OR "fecundity"
- *[SOCIAL] "*childlessness" OR "involuntary childlessness" OR "social infertility" OR "fertility intentions" OR "reproductive decision-making"

The resulting database of publications (101,645 clinical publications and 3,646 social publications) is compared across five dimensions: growth over time, metadata completeness, open access status, output type, and institutional provenance.

## 3. Key Findings

Three key findings emerge from the analysis. First, the social framing is harder to surface at scale as multiple independent indicators point and although identifier gaps are modest, these sit on top of a 28-fold difference in absolute volumn. Second, "best-indexed" is not "best evidence": what the data reveal is machine-legibility attrition, not quality attrition. The papers that disappear from automated pipelines are not worse; they are simply less conformant with the identifier and format infrastructure through which retrieval works. Third, clinical terms dominate not only academic output but policy discourse itself: the data shows more policy documents associated with the clinical framing, confirming that the asymmetry extends well beyond academia.

### Corpus Growth (1970–2025)

Fig 1 charts publication counts by year for both framings since 1970. Clinical output exceeds social in every year. The more significant observation is structural: the queryable record is weighted toward an older, established clinical framing, while the social framing represents the growth edge of the literature but remains marginal in absolute scale. The question this raises for AI evidence synthesis is whether a pipeline built on this record would reproduce the asymmetry, or amplify it, given that retrieval systems weight identifiers and citation density in ways that compound the underlying imbalance.

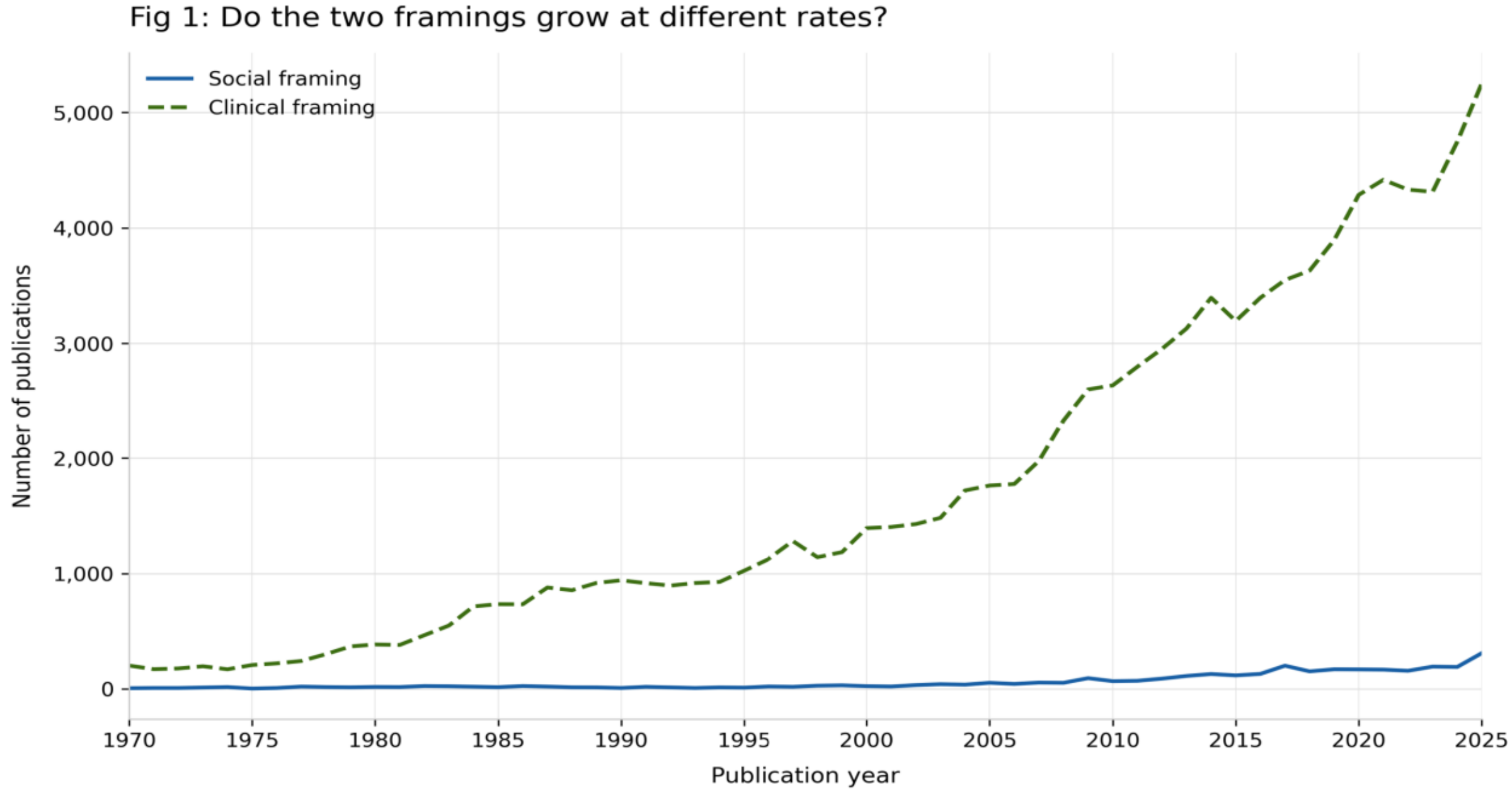

*Fig 1: Do the two framings grow at different rates? Publication counts by year, 1970-2025.*

## Metadata Completeness

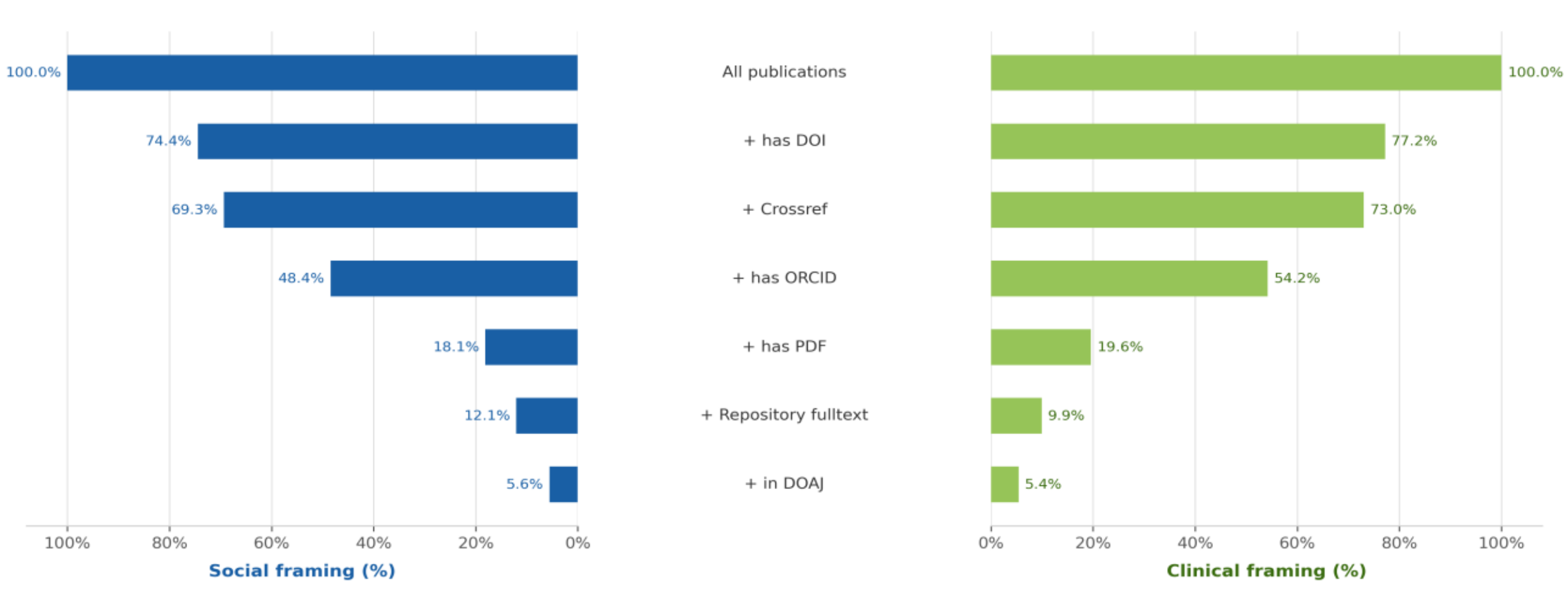

*Fig 2: How much of each corpus survives successive metadata filters?*

Fig 2 examines the metadata infrastructure layer directly, comparing the two bodies of literature across six key facets: DOI coverage, Crossref indexing, ORCID coverage, PDF availability, repository full-text and DOAJ listing. Clinical literature leads on the identifiers that retrieval systems prioritise most heavily. ORCID coverage is 67.6% versus 58.3%. Cumulative survival through all filters is nonetheless near-equal (5.6% social vs 5.4% clinical; 204 vs 5,489 works), so the social framing's disadvantage is absolute scale and identifier mix rather than proportional attrition. Repository full-text inverts: 30.1% social versus 22.6% clinical, suggesting social scholarship compensates for weaker formal infrastructure through self-archiving. The conclusion is that the social framing is less machine-legible on precisely the identifiers that automated retrieval systems prioritise.

## Output Types

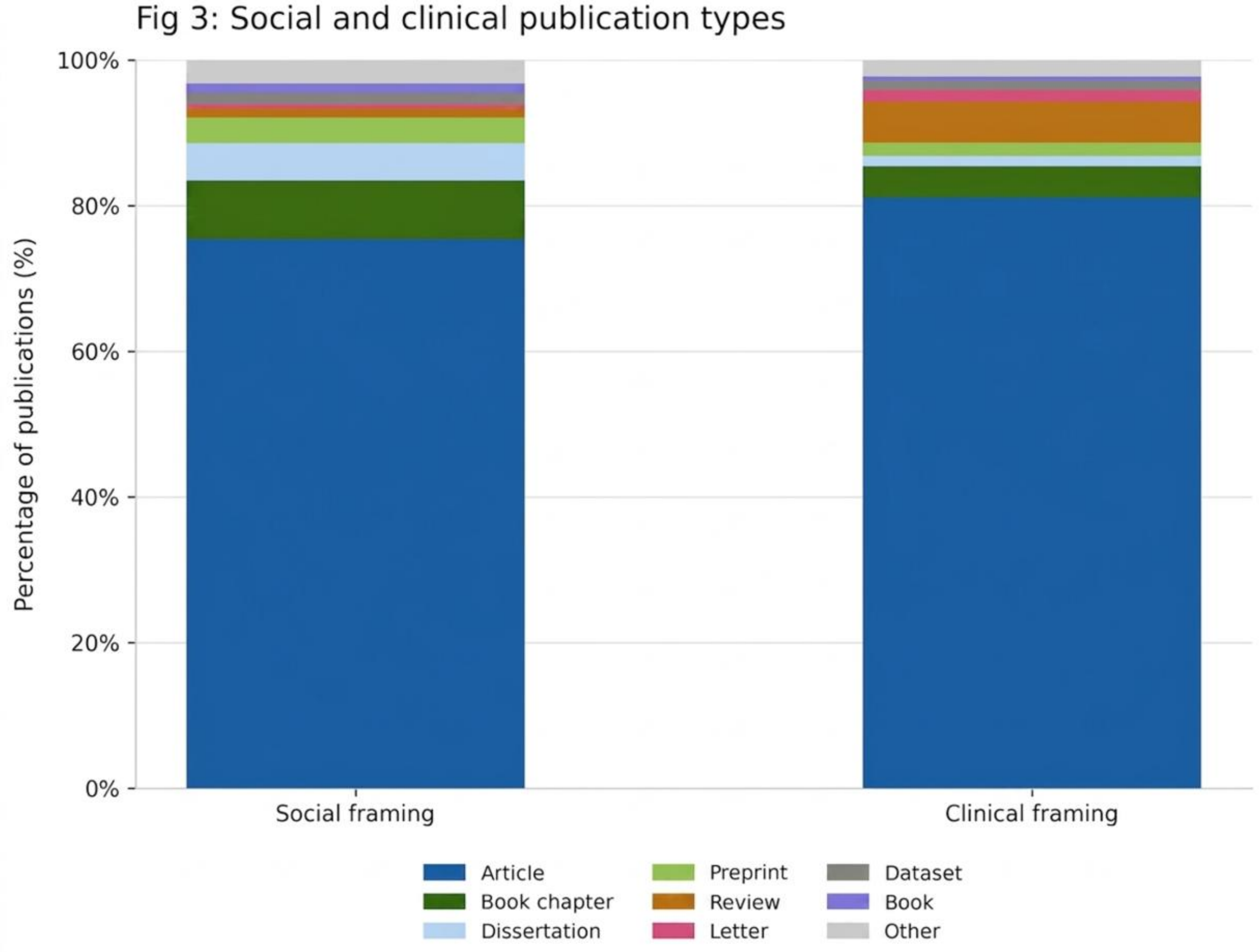

*Fig 3: Social and clinical publication types. Stacked bar chart of output type by framing.*

Even though both literatures are roughly equally open, the social framing is less machine-legible on the identifiers that retrieval systems prioritise. The gap is in indexing depth and format conformance, not in access. Fig 3 shows the structural source of the legibility gap. The social framing is roughly twice as reliant on books, chapters, and dissertations (15% vs 6%); clinical publishing concentrates in articles and reviews (~87%), the formats indexing infrastructure and AI pipelines are built to read. But format diversity carries a cost: qualitative and interpretive evidence in books and dissertations is systematically underread by article-centric synthesis pipelines. Dissertations in particular often lack DOIs, Crossref records, and structured metadata, and are frequently excluded from systematic review pipelines by design.

## Institutional Provenance

Fig 4 maps the institutional origins of both corpora. Healthcare institutions account for 27.6% of clinical authorships but only 10.1% of social. Social scholarship is university-dominated: education institutions represent 64.7% of social affiliations versus 49.7% of clinical.

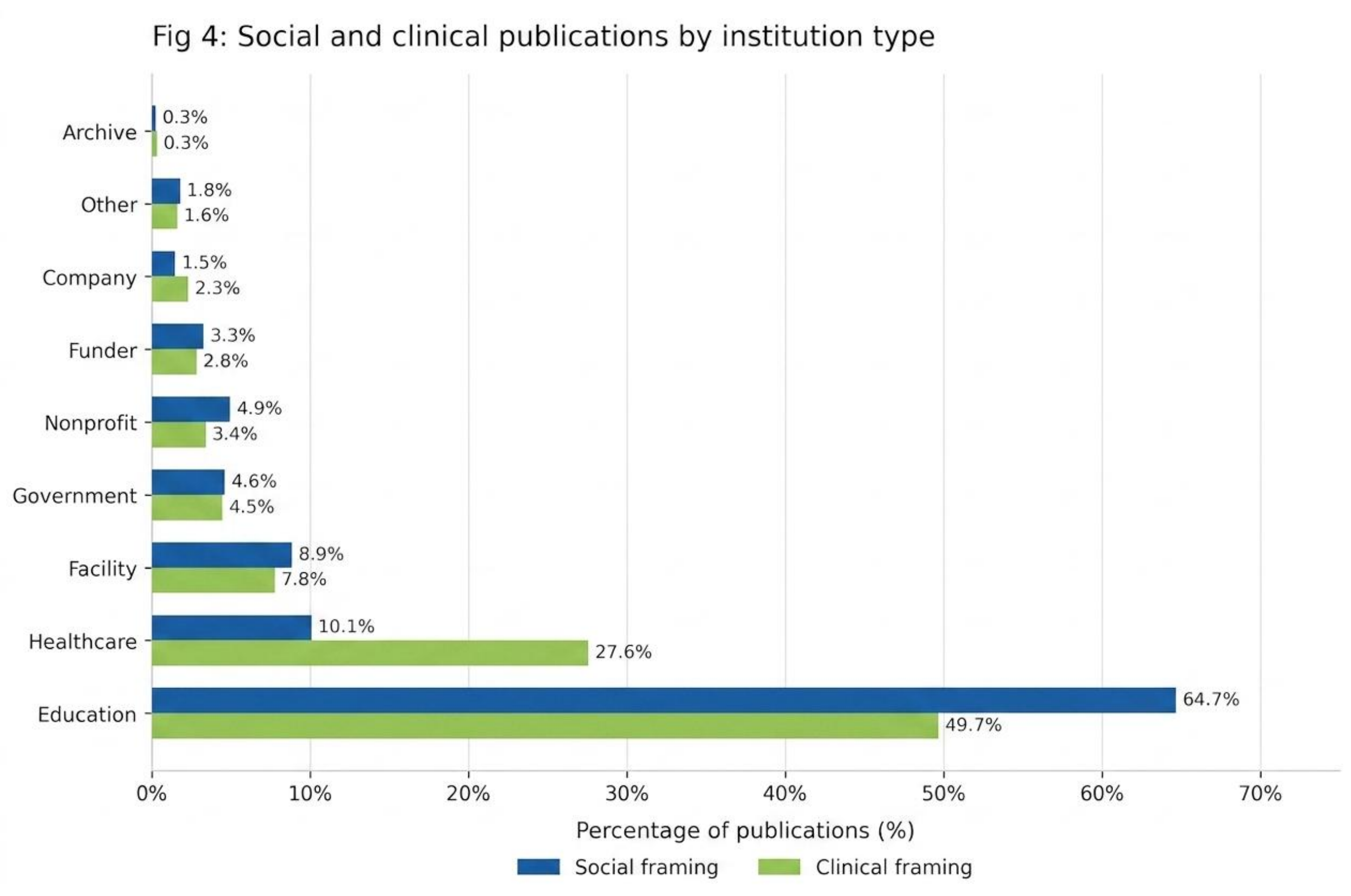

*Fig 4: Social and clinical publications by institution type. Horizontal grouped bar chart of authorship institution type by framing.*

This distribution explains the mechanism behind the results in Fig 2. Clinical scholarship originates disproportionately in healthcare settings, such as hospitals, medical centres, research clinics, where biomedical metadata infrastructure is embedded in daily workflows: PubMed submission, ORCID registration, and DOI assignment are routine. University-based social science operates in a different environment, where these systems are less consistently adopted. Institutional provenance, in other words, is a strong predictor of machine-legibility.

## 4. Conclusions

The imbalance documented across our analysis establishes the precondition for a hidden policy filter. Whether that filter actively distorts the outputs of AI evidence-synthesis tools and by how much, is the next empirical question. This requires a tool-output test, more data sources and replication across further contested topics.

We argue that evidence infrastructure quietly tilts the answer space on fertility decline toward clinical and ART interventions and away from the social dimensions, such as housing, care, economic security, and reproductive autonomy. As automated synthesis tools become more prevalent in policy briefing and evidence review, they will systematically under-read the social framing. The policy responses they surface will then also struggle to see the structural causes – not because the evidence does not exist, but because it has not been made legible to the machines doing the reading.

The governance implication follows directly. Discoverability infrastructure should be treated as a governance variable for AI evidence synthesis, subject to audit and investment alongside models, prompts, and inclusion criteria. Metadata standards and interoperability are not technical housekeeping, but the mechanism through which the balance of power in evidence-based policymaking is quietly negotiated. Making that mechanism visible is the first step toward governing it.